\documentclass[prd,reprint,aps,amsmath,amssymb,showpacs,nofootinbib]{revtex4-1}
\usepackage[utf8x]{inputenc}
\usepackage{graphicx}
\usepackage{enumitem}
\usepackage{slashed}
\usepackage{url}
\usepackage[pdftex]{hyperref}
\usepackage[T1]{fontenc}
\usepackage{bm}
\usepackage{color}

\newcommand{\bra}[1]{\langle #1|}
\newcommand{\ket}[1]{|#1 \rangle}
\newcommand{\braket}[2]{\langle #1|#2 \rangle}
\newcommand{\fslash}[1]{\!\not\!\!{#1}}
\newcommand{\MeV}{\rm MeV}
\newcommand{\GeV}{\rm GeV}
\newcommand{\TeV}{\rm TeV}
\newcommand{\geff}{g_{\ast}}
\begin{document}

\title{Baryogenesis via Mesino Oscillations}
\author{Akshay Ghalsasi}
\email{aghalsa2@uw.edu}
\affiliation{Department of Physics, University of Washington, Seattle, WA 98195, USA}

\author{David McKeen}
\email{dmckeen@uw.edu}
\affiliation{Department of Physics, University of Washington, Seattle, WA 98195, USA}

\author{Ann E. Nelson}
\email{aenelson@uw.edu}
\affiliation{Department of Physics, University of Washington, Seattle, WA 98195, USA}                             
\date{\today}

\begin{abstract}
We propose a new mechanism for baryogenesis at the 1-200 MeV scale. Enhancement of CP violation takes place via interference between oscillations and decays of {\it mesinos}---bound states of a scalar quark and antiquark and their CP conjugates. We present the mechanism in a simplified model with four new fundamental particles, with masses between 300 GeV and 10 TeV, and show that some of the experimentally allowed parameter space can give the observed baryon-to-entropy ratio. 
\end{abstract}

\maketitle

\section{Introduction}
{\it Baryogenesis}, the explanation for the asymmetry between matter and antimatter in our universe, is a profound puzzle for particle physics, as well as one of the strongest motivations for extending the standard model (SM). The first proposed solution, by Sakharov in 1967, laid out three conditions for a successful resolution to the puzzle~\cite{Sakharov:1967dj}. The first condition, C and CP violation, is satisfied by a CP-violating phase in the CKM matrix of the SM, however the effects of this phase are too suppressed in the early universe  to produce a sufficient asymmetry. The second condition, violation of baryon number, is satisfied by anomalous electroweak processes which are sufficiently fast at high temperature to produce baryon number~\cite{Kuzmin:1985mm,*Arnold:1987mh,*Arnold:1987zg,*Arnold:1996dy,*Moore:2000mx}, but only if the third condition, departure from thermal equilibrium, is satisfied. The minimal SM with a Higgs boson at 125 GeV does not have any phase transition or sufficiently long-lived heavy particles to produce a sufficient departure from thermal equilibrium for baryogenesis~\cite{Arnold:1992rz,*Dine:1992wr,*Buchmuller:1993bq,*Farakos:1994kx,*Kajantie:1995kf,*Csikor:1998eu}.

In most supersymmetric extensions of the standard model, cosmology theory favors a low reheat scale after inflation~\cite{Pagels:1981ke,*Weinberg:1982zq,*Ellis:1986zt,*Ellis:1990nb,*Moroi:1993mb,*Kallosh:1999jj,*Bolz:2000fu,Banks:1993en}.  Furthermore, in many theories, the absence of observed isocurvature perturbations favors a low inflation scale~\cite{Banks:1993en,Linde:1984ti,*Turner:1989vc,*Linde:1991km,*Lyth:1992tx,*Fox:2004kb,*Sikivie:2006ni}. A low inflation or reheat scale would imply that the baryogenesis scale must also be low, in some cases as low as an MeV. Although   proposals for baryogenesis at scales as low as an MeV exist~\cite{Claudson:1983js,*Dimopoulos:1987rk}, the time scale at this epoch is relatively long compared to typical particle physics scales, making departure from thermal equilibrium difficult to arrange. Models which do depart from thermal equilibrium at these low scales then produce additional entropy which   dilutes the baryon abundance.  In addition, CP violation at these low scales, if large enough to produce  enough baryon number  to survive the dilution,  can easily lead to particle electric  dipole moments which are in conflict with experimental bounds.  Finally, low energy baryon number violation can lead to rare decays in conflict with observation. It is therefore difficult to produce enough baryons at low energy while satisfying experimental constraints, motivating the exploration of new mechanisms for low scale baryogenesis, particularly ones which enhance the quantum mechanical interference which is necessary for CP violation.

In the SM  the CP-violating phase is unphysical if any of the small mixing angles or like-sign quark mass differences vanish. A reparameterization-invariant measure of CP violation~\cite{Jarlskog:1985ht} in the SM is thus very tiny. Nevertheless, the SM does provide a wealth of large CP-violating asymmetries in the decays of oscillating neutral mesons, illustrating the efficiency of oscillations for enhancing the effects of  CP-violating phases.   A proposal~\cite{Farrar:1993sp} that a similar enhancement of  CP violation could occur during an electroweak phase transition was shown not to work due to the too rapid rate of thermal decoherence~\cite{Gavela:1993ts,*Huet:1994jb}.

In this paper we present a new baryogenesis mechanism. We  show how in an extension of the SM, the particle-antiparticle oscillations of {\it mesinos}~\cite{Sarid:1999zx,Berger:2012mm} (bound states of a fermion quark and a scalar antiquark or vice versa),  when combined with baryon-violating scalar  decays, can give rise to  baryogenesis. The scalar quark, whose mass is of order a TeV,  is produced by the out of equilibrium decays of a long lived heavy neutral fermion, below  the QCD hadronization scale but before  nucleosynthesis. Both the oscillation rate and the decay rate of the mesinos are rapid when compared with the decoherence time, allowing   quantum mechanical interference between particle and anti particle decays to produce substantial CP violation.  This mechanism for producing baryons is amusingly similar to the mechanism for CP-violating production of charged leptons in the neutral kaon system.  We will illustrate the mechanism in a simplified  model which  introduces a minimal number of new particles beyond the SM.  Our mechanism and the new particles could be part of  a supersymmetric extension of the SM with baryon-number--violating R-parity violation.

The organization of the paper is as follows. In Sec.~\ref{sec:model} we introduce the model, describing the CP-violating oscillations of the mesino. Section~\ref{sec:exp} details experimental constraints on the model, from production at colliders and precision measurements of rare processes. In Sec.~\ref{sec:cosmology} we describe the cosmology of the model, estimating the parameters needed to explain the observed baryon asymmetry. We conclude in Sec.~\ref{sec:conclusions}.

\section{The Model and Some Phenomena}
\label{sec:model}
The ingredients we need for our model are a complex scalar $\phi$ and $n$ Weyl fermions $N_i$. $\phi$ is a color triplet, SU$(2)_L$ singlet, and carries hypercharge $-1/3$. Each $N_i$ is taken to be a singlet under the SM gauge group. The relevant interactions of these are
\begin{equation}
{\cal L}\supset y_{ij}\phi\, \bar{d_{i}}N_{j} -\frac12 m_{Nij}N_{i}N_{j} + \alpha_{ij}\phi^*\bar{d_{i}}\bar{u_{j}} + {\rm c.c.} 
\label{eq:lag}
\end{equation}
$\bar u_i$ and $\bar d_i$ are the  left handed up and down-type singlet antiquarks, respectively, with the subscript labeling the generation. Another possible baryon-number--violating  interaction of $\phi$, $\phi\, q_i q_j$, where the $q_i$ are quark doublets, is neglected because we are considering only the interactions that could come from baryon-number--violating interactions in a supersymmetric theory where $\phi$ is a superpartner of  a down-type singlet quark.  By rotating and rephasing the $N_i$ we can make the singlet mass matrix $m_N$ real and diagonal. $\alpha_{ij}$ is a $3\times 3$   matrix containing nine complex parameters, with seven phases removable by reparameterizations. The remaining phases in this matrix do not play a role in our baryogenesis mechanism and will be ignored for the rest of the paper. Having exhausted our freedom to rephase the fields,  each of the elements in the $3\times n$ matrix $y_{ij}$ contains a physical, CP-violating imaginary component.

We will see that the simplest version of the model involves three new Weyl fermions, i.e. $n=3$, which we label in order of their masses, $m_{N_1}<m_{N_2}<m_{N_3}$. The colored scalar is produced in the early Universe by late, out-of-equilibrium decays of $N_3$. The scalar forms color-singlet bound states with SM quarks, termed mesinos, which undergo particle-antiparticle oscillations and decay. The two lighter singlets provide common intermediate on- and off-shell states for the mesino and antimesino that allow for them to violate CP (and baryon number) in the interference between decays with and without mixing, sourcing the baryon asymmetry of the Universe.

In the following sections  we discuss mesino-antimesino oscillation in this model and the constraints from collider experiments and precision measurements.

\subsection{Mesino oscillations and decay}
At temperatures below the QCD hadronization scale, $T_c\simeq 200~\MeV$, if the colored scalars are sufficiently long-lived, they will bind with light SM quarks to form mesinos and antimesinos. In analogy with the naming convention for mesons, we will refer to the (electrically neutral) mesino containing $\phi^\ast$ and $q=d,s,b$ as $\Phi_q$ with its antiparticle $\bar\Phi_q$ containing $\phi$ and $\bar q=\bar d,\bar s,\bar b$.

$\Phi_q$ and $\bar\Phi_q$ form a pseudo-Dirac fermion and, as discussed in Ref.~\cite{Ipek:2014moa}, the  system can be described using a two-state Hamiltonian containing dispersive and absorptive parts, just like the case of the neutral mesons,
\begin{equation}
{\bm H}={\bm M}-\frac{i}{2}{\bm \Gamma}.
\label{eq:H}
\end{equation}
The eigenstates of $\bm H$, with eigenvalues $\omega_{L,H}$, can be given in terms of the flavor eigenstates $\ket{\Phi_q}$ and $\ket{\bar\Phi_q}$,
\begin{equation}
\ket{\Phi_{L,H}}=p\ket{\Phi_q}\pm q\ket{\bar\Phi_q},
\end{equation}
with $L$ and $H$ referring to light and heavy. The complex numbers $p$ and $q$ are related by
\begin{equation}
\left(\frac qp\right)^2=\frac{{\bm M}_{12}^\ast-\left(i/2\right){\bm \Gamma}_{12}^\ast}{{\bm M}_{12}-\left(i/2\right){\bm \Gamma}_{12}}.
\end{equation}
The mass and width differences between $\ket{\Phi_{L}}$ and $\ket{\Phi_{H}}$ are
\begin{align}
\Delta m&=m_H-m_L={\rm Re}\left(\omega_H-\omega_L\right),
\\
\Delta \Gamma&=\Gamma_H-\Gamma_L=-2{\rm Im}\left(\omega_H-\omega_L\right),
\end{align}
with
\begin{equation}
\omega_H-\omega_L=2\sqrt{\left({\bm M}_{12}-\frac i2{\bm \Gamma}_{12}\right)\left({\bm M}_{12}^\ast-\frac i2{\bm \Gamma}_{12}^\ast\right)}.
\end{equation}
In the scenario we will study, the mass difference is small compared to the masses of the heavy and light eigenstates,
\begin{align}
m_H\simeq m_L\simeq \frac{m_H+m_L}{2}\equiv m_{\Phi_q}.
\end{align}
Furthermore, the mass of the mesino is mostly supplied by the $\phi$, $m_{\Phi_q}\simeq m_\phi$. As in the case of meson oscillations, it is often useful to introduce a dimensionless parameter to characterize the oscillation rate,
\begin{align}
x\equiv \frac{\Delta m}{\Gamma},
\end{align}
where $\Gamma=(\Gamma_H+\Gamma_L)/2$ is the average lifetime of the states. $x\gg 1$ indicates that the mesino system oscillates rapidly before decaying while $x\ll 1$ means that the mesino typically decays before oscillating much. 

We use $\ket{\Phi_q\left(t\right)}$ to label a state at time $t$ that began at $t=0$ as a pure mesino and $\ket{\bar\Phi_q\left(t\right)}$ to label one that began as a pure antimesino. These evolve in time according to
\begin{align}
\ket{\Phi_q\left(t\right)}&=g_+\left(t\right)\ket{\Phi_q}-\frac qp g_-\left(t\right)\ket{\bar\Phi_q},
\\
\ket{\bar\Phi_q\left(t\right)}&=g_+\left(t\right)\ket{\bar\Phi_q}-\frac pq g_-\left(t\right)\ket{\Phi_q},
\end{align}
with
\begin{align}
g_\pm\left(t\right)=\frac12\left(e^{-im_Ht-\frac12\Gamma_Ht}\pm e^{-im_Lt-\frac12\Gamma_Lt}\right).
\end{align}

The baryon-number--violating operator $\phi^*\bar{d_{i}}\bar{u_{j}}$ and its conjugate in Eq.~(\ref{eq:lag}) allow for the mesino and antimesino flavor eigenstates to decay to collections of hadrons with baryon number $B=+1$ and $B=-1$, respectively, with amplitudes related by complex conjugation. In terms of the time-independent amplitude for a mesino to decay to a state with $B=+1$, $\cal M$, the time-dependent amplitudes for initial mesino or antimesino states to decay to $B=\pm 1$ (denoted $B$ and $\bar B$) are
\begin{equation}
\begin{aligned}
\braket{B}{\Phi_q\left(t\right)}&=g_+\left(t\right){\cal M},
\\
\braket{\bar B}{\Phi_q\left(t\right)}&=-\frac qp g_-\left(t\right){\cal M}^\ast,
\\
\braket{B}{\bar\Phi_q\left(t\right)}&=-\frac pq g_-\left(t\right){\cal M},
\\
\braket{\bar B}{\bar\Phi_q\left(t\right)}&=g_+\left(t\right){\cal M}^\ast.
\end{aligned}
\end{equation}
Squaring these and integrating over $t$, we can find the probability that an initial mesino state decays to $B$ and $\bar B$,
\begin{align}
P_{\Phi_q\to B}&\propto\int_0^\infty dt\left|\braket{B}{\Phi_q\left(t\right)}\right|^2,
\\
P_{\Phi_q\to \bar B}&\propto\int_0^\infty dt\left|\braket{\bar B}{\Phi_q\left(t\right)}\right|^2,
\end{align}
and similarly for an initial antimesino. Using these expressions, we write down the baryon asymmetry per mesino-antimesino pair,
\begin{align}
\epsilon_B=A_B\times{\rm Br}_{\Phi_q\to B},
\label{eq:eps}
\end{align}
where
\begin{equation}
A_B=\frac{P_{\Phi_q\to B}-P_{\Phi_q\to \bar B}+P_{\bar\Phi_q\to B}-P_{\bar\Phi_q\to \bar B}}{P_{\Phi_q\to B}+P_{\Phi_q\to \bar B}+P_{\bar\Phi_q\to B}+P_{\bar\Phi_q\to \bar B}}.
\end{equation}
and ${\rm Br}_{\Phi_q\to B}=\Gamma_{\Phi_q\to B}/\Gamma$ is the (time-independent) branching ratio of a mesino flavor eigenstate to decay into a state with $B=+1$. In terms of the elements of $\bm H$, $A_B$ is
\begin{equation}
A_B=\frac{2{\rm Im} {\bm M}_{12}^\ast{\bm \Gamma}_{12}}{\Gamma^2+4\left|{\bm M}_{12}\right|^2}.
\label{eq:A_B}
\end{equation}

To get a sense of how large the asymmetry $\epsilon_B$ can be, we estimate the magnitudes of the elements that appear in it below.

\subsubsection{Estimating $\mathbf M_{12}$}
\label{sec:M12}
\begin{figure}
\includegraphics[width=\linewidth]{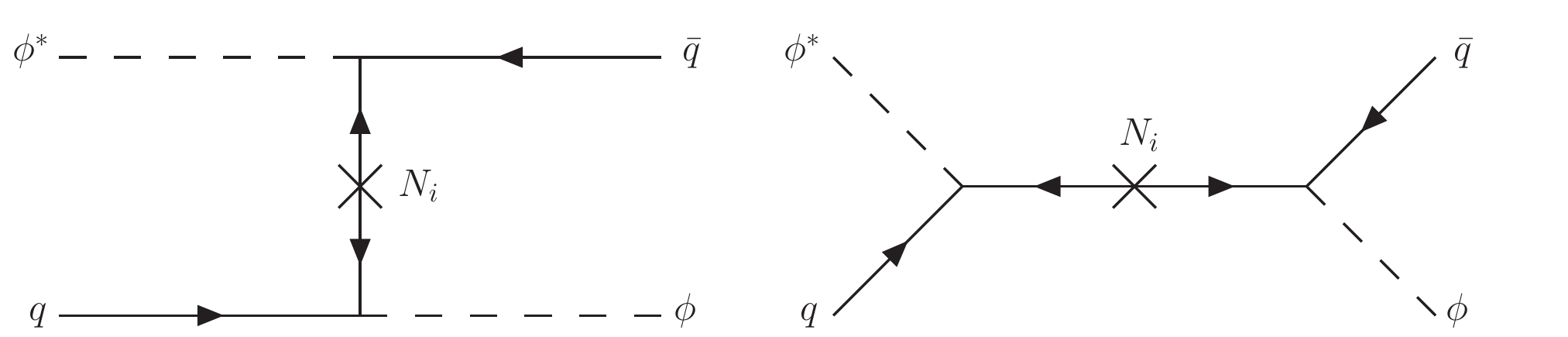}
\caption{Tree level contributions to ${\bm M}_{12}$.}\label{fig:M12}
\end{figure}

The leading contributions to ${\bm M}_{12}$ in the $\Phi_q$-$\bar\Phi_q$ system arise from the diagrams in Fig.~\ref{fig:M12}, resulting in~\cite{Berger:2012mm,Sarid:1999zx}
\begin{equation}
{\bm M}_{12}  = \sum_i {\bm M}_{12}\left(N_i\right)=\frac{2f^{2}_{{\Phi_q}}}{3} \sum_j  \frac{y^{2}_{qi} m_{N_i}}{m^{2}_{N_i}-m^{2}_{\Phi_{q}}},
\end{equation}
where we have written the contribution from each $N_i$ as ${\bm M}_{12}\left(N_i\right)$.

The mesino decay constant $f_{{\Phi_q}}$ can be related to the $B$ meson decay constant~\cite{Manohar:2000dt},
\begin{equation}
f_{{\Phi_q}}= f_{B_q}\sqrt{\frac{m_b}{m_\phi}}\left(\frac{\alpha_s\left(m_b\right)}{\alpha_s\left(m_t\right)}\right)^{6/23}\left(\frac{\alpha_s\left(m_t\right)}{\alpha_s\left(m_\phi\right)}\right)^{6/21}.
\end{equation}
The cosmology of the model will motivate focusing on the strange mesino; to estimate its decay constant, we use $f_{B_s}=225~\MeV$~\cite{Na:2012kp}, $\overline{\rm MS}$ quark masses, $m_b=4.18~\GeV$, $m_t=160~\GeV$~\cite{Agashe:2014kda}, and leading order QCD running (the contribution of which is negligible above the top mass in this case), and find
\begin{align}
f_{{\Phi_s}}&\simeq21.5~{\MeV}\sqrt{\frac{650~\GeV}{m_\phi}}.
\label{eq:decayconst}
\end{align}
In what follows, we will take $q=s$, specifying to the case of the strange mesino, $\Phi_s$.

\subsubsection{Estimating $\Gamma$ and $\mathbf\Gamma_{12}$}
\label{sec:Gamma}
As mentioned above, the operator $\phi^*\bar{d_{i}}\bar{u_{j}}$ in Eq.~(\ref{eq:lag}) allows for the mesino to decay to a collection of hadrons with $B=+1$ with a rate
\begin{equation}
\Gamma_{\Phi_s\to B}=\frac{1}{16\pi}\sum_{i,j}\left|\alpha_{ij}\right|^2m_{\Phi_s}= \frac{\alpha_B^2}{16\pi}m_{\Phi_s},
\end{equation}
where we assume that we can ignore the masses of the hadrons in the final state in comparison to the mesino's and we have defined $\alpha_B^2\equiv\sum_{i,j}\left|\alpha_{ij}\right|^{2}$.

In general, the mesinos can also decay to a singlet plus hadrons, $\Phi_q\to N_j+h$ where $h$ labels the hadronic state, through the operator $\phi \bar d_i N_j$ with a rate proportional to $\left|y_{ij}\right|^2$. This is kinematically allowed for $m_{\Phi_q}>m_{N_j}+m_{h}$. If $h$ is self-conjugate under C ($h=\pi^0,\pi^+\pi^-,\rho,\dots$) then this is a final state for antimesino decay as well, leading to a nonzero value of $\mathbf\Gamma_{12}$ and CP violation through Eq.~(\ref{eq:A_B}). This CP asymmetry is suppressed when $\left|{\bm \Gamma}_{12}\right|/\Gamma$ is small; for this reason we take only $N_1$ to be kinematically allowed in the decay of the strange mesino, and set its mass close enough to the mesino's so that single hadron final states dominate the partial width. In this case $\left|{\bm \Gamma}_{12}\right|\sim\Gamma_{\Phi_s\to N_1}$ and $\left|{\bm \Gamma}_{12}\right|/\Gamma$ can potentially be unsuppressed. 

\begin{figure}
\includegraphics[width=\linewidth]{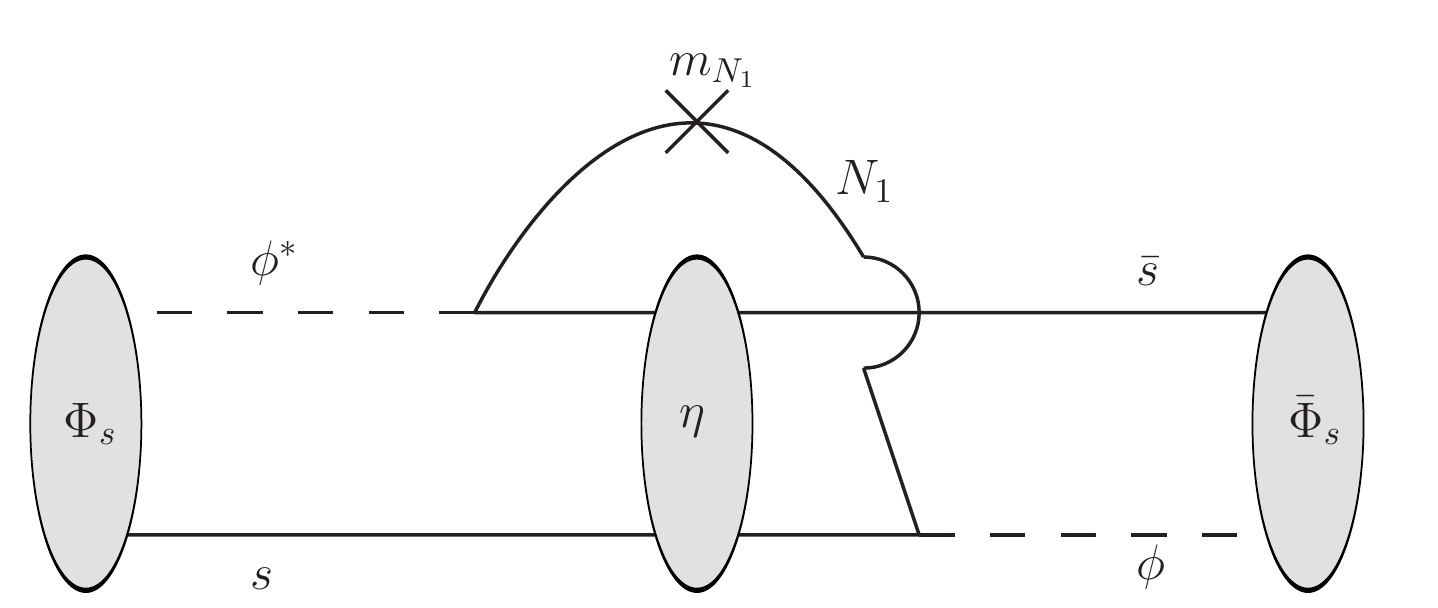}
\caption{2-body contribution to $\mathbf\Gamma_{12}$ from $\Phi_s,\bar\Phi_s\to N_1+\eta$ where $\eta$ is the lightest meson containing $s\bar s$.}\label{fig:Gamma12}
\end{figure}

Assuming that $m_{\Phi_s}\simeq m_{N_i}$, we can approximate $\mathbf\Gamma_{12}$ as being dominated by the decays $\Phi_s,\bar\Phi_s\to N_1+\eta$, where $\eta$ is the lightest pseudoscalar meson composed of $s\bar s$, as shown diagrammatically in Fig.~\ref{fig:Gamma12}. This results in
\begin{align}
\mathbf\Gamma_{12}&\simeq \mathbf\Gamma_{12}^{\rm 2-body}
\\
&=\frac{y_{s1}^2m_{N_1}}{16\pi}\left|F\left(m_{N_1}^2\right)\right|^2\lambda^{1/2}\left(1,\frac{m_{N_1}^2}{m_{\Phi_q}^2},\frac{m_P^2}{m_{\Phi_q}^2}\right),
\end{align}
where $\lambda\left(a,b,c\right)=a^2+b^2+c^2-2ab-2ac-2bc$ and $F\left(q^2\right)$ is the $\Phi_s\to P$ transition form factor at momentum transfer $q$.
We provide estimates of the form factor $F\left(m_{N_1}^2\right)$ in Appendix~\ref{sec:Gamma12appendix}. For $m_{\Phi_s}\sim1~\TeV$ with a splitting $m_{\Phi_s}-m_{N_1}\sim 1~\GeV$, the form factor is ${\cal O}\left(10^{-3}-10^{-2}\right)$. Defining the mass splitting, $\Delta m_{\phi N_i}\equiv\left| m_{\Phi_s}-m_{N_i}\right|$ and assuming that they are small compared to the mesino mass, the rate to decay to the lightest singlet and the off-diagonal term in the absorptive part of the Hamiltonian are given by
\begin{equation}
\begin{aligned}
\Gamma_{\Phi_q\to N_1}&\simeq\left|{\bm \Gamma}_{12}\right|\simeq 4\times10^{-8}~{\GeV}\left|\frac{y_{s1}}{0.1}\right|^2
\\
&\quad\quad\quad\quad\quad\times\left|\frac{F\left(m_{N_1}^2\right)}{10^{-2}}\right|^2\left(\frac{\Delta m_{\phi N_1}}{1~\GeV}\right),
\label{eq:Gamma12}
\end{aligned}
\end{equation}
for a scalar of mass $\sim1~\TeV$.

For the mesino to hadronize before decaying, the decay rate has to be slow compared to the timescale for hadronization, $\Gamma\ll 1/t_{\rm had}\sim\Lambda_{\rm QCD}\sim 200~\MeV$. The partial width to $N_1$ satisfies this condition for any set of parameters we consider. Requiring that the partial width to baryons is smaller than $\Lambda_{\rm QCD}$ limits $\alpha_B\lesssim 0.1\sqrt{650~{\GeV}/m_\phi}$.

\subsubsection{Simplifying the asymmetry}
In the limit of small mass splittings between $\Phi_s$ and $N_i$, $\Delta m_{\phi N_i}\ll m_\phi$, then the contribution to $\bm M_{12}$ from $N_i$ is approximately
\begin{equation}
\begin{aligned}
\left|{\bm M}_{12}\left(N_i\right)\right|&\simeq 1.54\times10^{-6}~{\GeV}\left|\frac{y_{si}}{0.1}\right|^2
\\
&\quad\quad\quad\times\left(\frac{650~\GeV}{m_\phi}\right)\left(\frac{1~\GeV}{\Delta m_{\phi N_i}}\right),
\label{eq:M12}
\end{aligned}
\end{equation}
choosing to normalize the mass of the scalar above experimental constraints from collider searches (see Sec.~\ref{sec:colliders} below). Comparing this with Eq.~(\ref{eq:Gamma12}), we expect that, barring extreme cancellations between contributions from the different singlets, $\left|{\bm M}_{12}\right|\gg {\bm \Gamma}_{12},\Gamma_{\Phi_q\to N_1}$. 

The asymmetry per mesino pair can be written simply in terms of the oscillation parameter $x$ in this limit. When $\left|{\bm \Gamma}_{12}\right|\ll\left|{\bm M}_{12}\right|$, $x\simeq 2\left|{\bm M}_{12}\right|/\Gamma$ and
\begin{equation}
\epsilon_B= \frac{x\, r \sin\beta}{1+x^2}\frac{\left|{\bm \Gamma}_{12}\right|}{\Gamma}{\rm Br}_{\Phi_q\to B}+{\cal O}\left(\left|\frac{{\bm \Gamma}_{12}}{{\bm M}_{12}}\right|\right),
\end{equation}
where
\begin{align}
r&\equiv\left|1-\frac{{\bm M}_{12}\left(N_1\right)}{{\bm M}_{12}}\right|,~0<r<1,
\end{align}
and the reparameterization-invariant, CP-violating phase is
\begin{align}
\beta&\equiv\arg\left\{\left[{\bm M}_{12}-{\bm M}_{12}\left(N_1\right)\right]^\ast{\bm \Gamma}_{12}\right\}.
\end{align}
$\beta$ is defined this way so as to be generically ${\cal O}\left(1\right)$, making use of the fact that ${\bm \Gamma}_{12}$ and ${\bm M}_{12}\left(N_1\right)$ have the same phase. If we also make use of the relation between the decay rate to a singlet and the off-diagonal term in $\bm\Gamma$, $\left|{\bm \Gamma}_{12}\right|\simeq\Gamma_{\Phi_q\to N_1}=\Gamma\,{\rm Br}_{\Phi_q\to\bar N_1}$ then
\begin{equation}
\epsilon_B\simeq\frac{x\, r \sin\beta}{1+x^2}{\rm Br}_{\Phi_q\to B}{\rm Br}_{\Phi_q\to N_1}.
\end{equation}

The question of whether $\left|{\bm M}_{12}\right|$ is large or small compared to the total mesino decay rate $\Gamma$ is determined by the partial width to antibaryons, $\Gamma_{\Phi_q\to B}$. If  $\alpha_B$ is much smaller than about $10^{-4}\left|y_{s1}\right|(650~{\GeV}/m_\phi)$, then $\left|{\bm M}_{12}\right|\gg \Gamma_{\Phi_q\to B}$ and $x\gg 1$. Then,
\begin{equation}
\epsilon_B\to\frac{r}{x} \sin\beta\,{\rm Br}_{\Phi_q\to B}{\rm Br}_{\Phi_q\to N_1}.
\label{eq:xlarge}
\end{equation}
In this case, by inspecting Eqs.~(\ref{eq:M12}) and (\ref{eq:Gamma12}), typically $x\lesssim 10^2$, depending on the hierarchy of $\left|{\bm \Gamma}_{12}\right|$ and $\Gamma_{\Phi_q\to B}$.  The asymmetry is suppressed by $x$ and a potentially small branching ratio. On the other hand, if  $\alpha_B \gg 10^{-4}\left|y_{s1}\right|(650~{\GeV}/m_\phi)$ then $\left|{\bm \Gamma}_{12}\right|\ll\left|{\bm M}_{12}\right|\ll \Gamma_{\Phi_q\to B}$ and $x\ll 1$. The asymmetry becomes
\begin{equation}
\epsilon_B\to x\,r\sin\beta\,{\rm Br}_{\Phi_q\to\bar N_1},
\label{eq:xsmall}
\end{equation}
suppressed by the small values of both $x$ and ${\rm Br}_{\Phi_q\to\bar N_1}$.

The asymmetry is typically largest when $\Gamma_{\Phi_q\to B}\simeq\left|{\bm \Gamma}_{12}\right|\ll\left|{\bm M}_{12}\right|$. In this situation Eq.~(\ref{eq:xlarge}) applies and $\epsilon_B\sim10^{-2} r\sin\beta$. We note that it is possible to achieve an asymmetry as large as $\epsilon_B=1/8$ if $\Gamma_{\Phi_q\to B}\simeq\left|{\bm \Gamma}_{12}\right|\simeq\left|{\bm M}_{12}\right|$. However, this requires large, unnatural cancellations (at the level of $10^{-2}$) between the contributions to ${\bm M}_{12}$.

\section{Experimental Constraints}
\label{sec:exp}
 
\subsection{Collider Constraints}
\label{sec:colliders}
In this model the colored scalar can be pair produced in $pp$ and $p\bar p$ collisions which then decay to multijet final states. If the dominant decay mode is through the coupling $\alpha_{ij}\phi^\ast\bar d_i \bar u_j+{\rm h.c.}$, then $\phi$ will decay to two hard jets, leading to a four-jet signature. If the decay of $\phi$ proceeds mostly through $N_1$, the situation is somewhat more nuanced as we describe below.

In proton-(anti)proton collisions, the $\phi\phi^\ast$ pair is produced in the lab frame with very little boost in the direction transverse to the beam. The $\phi$  decays to a very soft light quark with a momentum of ${\cal O}\left(\Delta m_{\phi N_1}\right)$ and $N_1$ that is also essentially at rest in the $\phi$ rest frame, because of the small mass splitting between the $\phi$ and $N_1$. The $N_1$ decays through an off-shell $\phi$ to three quarks with a rate
\begin{align}
\frac{d\Gamma_{N_1}}{dp}&\simeq\sum_{i}\frac{\left|y_{i1}\right|^2\alpha_B^2}{512\pi^3}\frac{p}{m_{N_1}}\left(\frac{m_{N_1}-2p}{p+\Delta m_{\phi N_1}}\right)^2,
\end{align}
ignoring the masses of the quarks. $p$ is the momentum of the quark produced in the $N_1\to\phi \bar d_i$ splitting in the $N_1$ rest frame. Because of the small mass splitting, the momentum of this quark is peaked at small values---in most decays $p\lesssim\sqrt{\Delta m_{\phi N_1} m_{N_1}}$. This means that, for a splitting $\Delta m_{\phi N_1} =1~\GeV$, $N_1$ masses below $\sim 1~\TeV$ result in one of the three quarks in $N_1$ decays having a transverse momentum in the lab frame of $p_T\lesssim30~\GeV$, making them hard to observe. Therefore, at colliders, $\phi$'s either decay directly to two hard jets or appear as two hard jets in their decays to $N_1$.

To use collider searches to set limits, we will focus on the situation where the scalars are produced and decay promptly. At the Large Hadron Collider (LHC) this requires decay lengths $c\tau_\phi,c\tau_{N_1}\lesssim1~\rm mm$. If the $\phi$'s decay directly to two quarks this requires $\alpha_B\gtrsim 10^{-7}\sqrt{650~{\GeV}/m_\phi}$. If, instead, $\phi$ decays proceed through $N_1$, then $(\sum_{i}\left|y_{i1}\right|^2)^{1/2}\gtrsim 10^{-4}$, for a form factor of $10^{-2}$ and a mass splitting of $1~\GeV$ and $(\sum_{i}\left|y_{i1}\right|^2)^{1/2}\alpha_B\gtrsim 10^{-6}\sqrt{650~{\GeV}/m_{N_1}}$. We will see in Sec.~\ref{sec:rare} that $\left|y_{b1}\right|$ is essentially unconstrained so this last constraint can be taken to be $\alpha_B\gtrsim 10^{-6}\sqrt{650~{\GeV}/m_{N_1}}$. The collider limits we will outline are relatively insensitive to the particular values of these couplings so long as they are large enough for prompt decays, given $\phi\phi^\ast$ production is dominated by QCD. We can also safely assume that these couplings are not so large as to make the width of $\phi$ large enough to impact the limits.

The requirement of prompt decays at hadron colliders is not a strict condition for the viability of the model, but merely simplifies our analysis later; a more robust lower bound, weaker on the couplings by ${\cal O}\left(10^6\right)$, comes from not spoiling Big Bang nucleosynthesis (BBN). Indeed, long-lived particles at colliders have been studied as a probe of baryogenesis, see, e.g., Ref.~\cite{Cui:2014twa}.
In the long-lived case, the lower limits on the particle masses can be quite strong, as high as $\sim1~\TeV$~\cite{Cui:2014twa}. 

A search for squarks decaying to a $b$ and an $s$ quark was undertaken at the LHC by the ATLAS collaboration using $17.4~{\rm fb}^{-1}$ of $8~\TeV$ $pp$ collisions~\cite{ATLAS-CONF-2015-026}. Squarks with masses between $100~\GeV$ and $310~\GeV$ were ruled out at the 95\%~confidence level (CL). Since they are both  color triplet scalars, the production cross section for a $\phi\phi^\ast$ pair is equal to that of a squark-antisquark pair. Thus, this exclusion would directly apply to our model if the leading $\alpha_{ij}$ were $\alpha_{31}$ or $\alpha_{32}$, and constrains $m_\phi>310~\GeV$.\footnote{A similar limit $\sim300~\GeV$ was derived in Ref.~\cite{Berger:2012mm} for mesinos oscillating and decaying to a top and a light quark, leading to a same-sign dilepton signal which was constrained by CMS data~\cite{Chatrchyan:2012paa}.} In addition, ATLAS has searched for the pair production of scalar gluons that each decay to two gluons using $4.6~{\rm fb}^{-1}$ collected at $7~\TeV$~\cite{ATLAS:2012ds}, finding no excess and setting an upper limit on the cross section for this process. This limit applies to $\phi$ in the case that it decays to a pair of light flavor quarks, which is the case if the $\alpha_{ij}$, $i,j=1,2$ couplings dominate. Using \texttt{NLL-fast}~\cite{Beenakker:1996ch,*Kulesza:2008jb,*Kulesza:2009kq,*Beenakker:2009ha,*Beenakker:2011fu} to calculate the cross section to produce a $\phi\phi^\ast$ pair at next-to-leading order with next-to-leading logarithmic corrections, this search sets the lower limit $m_\phi>275~\GeV$ at 95\% CL.

The CMS collaboration performed a search with $19.4~{\rm fb}^{-1}$ of $8~\TeV$ data, looking for the pair production of resonances that decay to two jets~\cite{Khachatryan:2014lpa}. Colored scalars decaying to a $b$ and a light flavor quark are ruled out at 95\% CL for $200~{\GeV}<m_\phi<385~{\GeV}$ and those decaying to two light quarks are excluded for $200~{\GeV}<m_\phi<350~{\GeV}$ at 95\% CL.

Although we expect the $\phi$ to decay to essentially two hard   jets, as explained above, if it decays through $N_1$, there could be apparent six-jet events from $\phi\phi^\ast$ production and decay. Searches for three-jet resonances~\cite{ATLAS:2012dp,*Aad:2015lea,*Chatrchyan:2013gia} generally have more reach than those for two-jet resonances because of the comparative ease of distinguishing this from backgrounds. In cases where these decays include heavy flavor, limits as high as $600~{\GeV}$ can be set.

In light of this discussion, we consider $600~\GeV$ as a conservative lower bound on the mass of $\phi$, deferring a complete study of the collider limits for future work. For this reason, when specifying $m_\phi$, we have taken $650~\GeV$ as a benchmark value.

The prospects for future searches for $\phi\to 2j$ are promising, with a study of the $14~\TeV$ LHC indicating that $m_\phi$ up to $750~\GeV$ could be probed with $300~{\rm fb}^{-1}$ and up to $1070~\GeV$ with $1000~{\rm fb}^{-1}$~\cite{Duggan:2013yna}. Such searches would push significantly into the region of parameter space compatible with an explanation of the baryon asymmetry of the Universe, as we will see in Sec.~\ref{sec:cosmology}.

\subsection{Constraints from Rare Processes}
\label{sec:rare}
 Our model possesses new interactions which violate CP, quark flavor, and baryon number, potentially leading to observable conflicts with Standard Model predictions. In examining the consequences, we note that  our model could be part of an R-parity--violating supersymmetric model~\cite{Fayet:1974pd,*Volkov:1973ix,*Hall:1983id,*Ross:1984yg,Zwirner:1984is,Barbieri:1985ty}, and we could use some of the existing studies of constraints on such models~\cite{Dawson:1985vr,*Dimopoulos:1988jw,*Carpenter:2006hs,*Evans:2012bf,*Brust:2011tb,*Giudice:2011ak,*Csaki:2011ge,*Bai:2013xla,*Allanach:2013qna,Ellis:1984gi,Barger:1989rk,Goity:1994dq,Barbier:2004ez}. The constraints on baryon-number--violating parameters in supersymmetric theories depend on    many   soft supersymmetry breaking parameters which play no role in baryogenesis. In this section we consider  the constraints on the parameters in our more minimal model.
Although baryon number is violated, given that that lepton number is conserved,\footnote{or conserved mod 2, if there are Majorana neutrino masses} and given that the only fermions lighter than the proton are leptons, the proton is stable. There are  stringent constraints on couplings involving the first generation  from neutron-antineutron oscillations, and on baryon-violating couplings of the strange quark, from heavy nuclei decays.

\begin{figure}
\includegraphics[width=\linewidth]{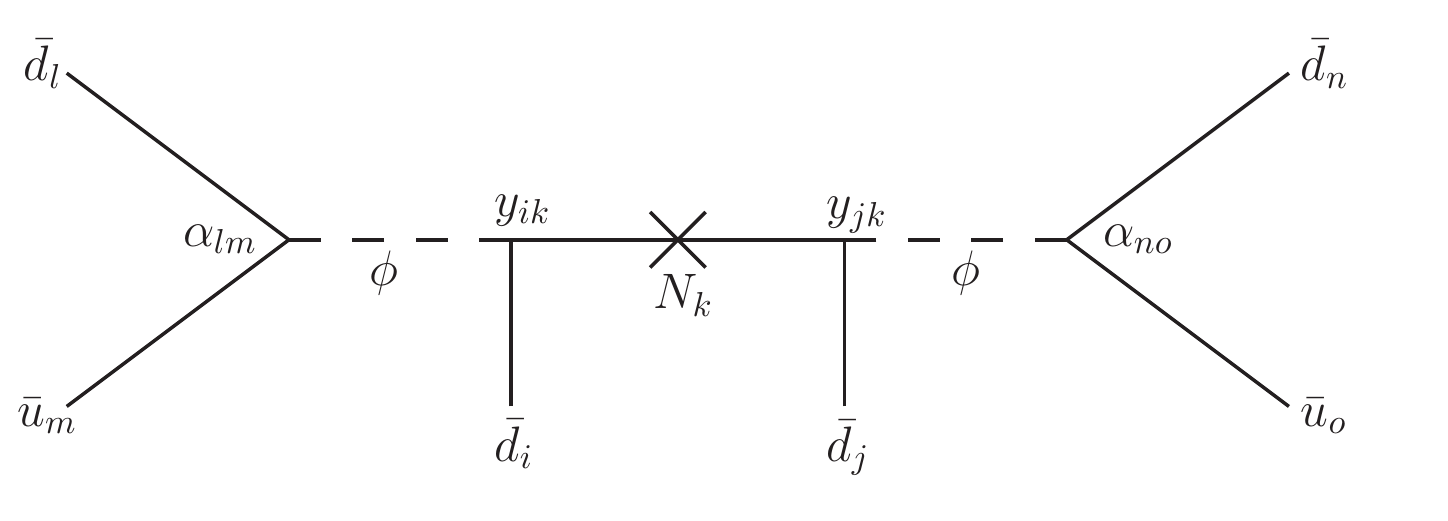}
\caption{Diagram responsible for the dimension-9 $\Delta  B=2$ operator in Eq.~(\ref{eq:nnbar}) that leads to neutron-antineutron oscillations.}\label{fig:nnbar}
\end{figure}

Dimension-9 six quark $\Delta B=2$ operators arise in the low energy effective theory from the diagram shown in Fig.~\ref{fig:nnbar}. The relevant operators are of the form
\begin{align}
   \left(\sum_k\frac{y_{ik}y_{jk}}{M_{N_k}}\right)\frac{\alpha_{lm}\alpha_{no}}{m_\phi^4} \bar{d}_i\bar{d}_j \bar{d}_l\bar{u}_m\bar{d}_n\bar{u}_o \ .\label{eq:nnbar}
\end{align}

Constraints on such operators with various color and spin combinations from neutron-antineutron oscillations have been considered in many previous works~\cite{Kuo:1980ew,*Rao:1982gt,*Dolgov:2006ay,*Babu:2013yww,Ozer:1982qh,Ellis:1984gi,*Barger:1989rk,Zwirner:1984is,Barbieri:1985ty,Abe:2011ky}. An accurate computation of the rate requires perturbative renormalization group scaling  and matching to lattice QCD matrix element computations of the various operators~\cite{Ozer:1982qh,
Buchoff:2012bm,*Phillips:2014fgb,*Buchoff:2015wwa,*Buchoff:2015qwa}, and currently still has uncertainties of order 1.  Currently the best constraint on the neutron oscillation rate is  $2.9 \times 10^{-33}~\GeV$~\cite{Abe:2011ky}. If we simply  estimate the matrix element of 
\begin{align}
\langle \bar{n} | \bar{d} \bar{d} \bar{d} \bar{d} \bar{u}\bar{u} |n\rangle\sim  {\cal O}(10^{-5}\ \GeV^6)\ ,
\end{align}
which is  consistent with    the lattice computations, we find
\begin{align} \left(\sum_k\frac{y_{dk}^2 }{M_{N_k}}\right)\frac{\alpha_{11}^2}{m_\phi^4} <  2.9\times 10^{-28}~\GeV^{-5}  \end{align}
Neglecting the very weakly coupled $N_3$, and using $M_{N_1}\sim M_{N_2}\sim m_\phi\sim 650\ GeV$ gives
\begin{align} \left(y_{d1}^2+y_{d2}^2\right)\alpha_{11}^2 < {\cal O} ( 10^{-14} ). \end{align}
The best constraints on operators involving strange quarks arise from limits on dinucleon decay into kaons~\cite{Barbieri:1985ty,Dimopoulos:1987rk,Goity:1994dq,Litos:2014fxa}. Using the rough estimates from Ref.~\cite{Goity:1994dq} for the nuclear matrix elements and the experimental  bound from Ref.~\cite{Litos:2014fxa} gives a bound on the coefficient of the operator $\bar{s} \bar{s} \bar{d} \bar{d} \bar{u}\bar{u}$ which, for $M_{N_1}\sim M_{N_2}\sim m_\phi\sim 650\ \GeV$, translates into   bounds on   combinations of couplings
\begin{align}\left(y_{s1}^2+y_{s2}^2\right)\alpha_{11}^2&< {\cal O} (  10^{-14} ),\\
\left(y_{d1}^2+y_{d2}^2\right)\alpha_{12}^2&< {\cal O} (  10^{-14} ),
\\
\left(y_{d1}y_{s1}+y_{d2} y_{s2}\right)\alpha_{12}\alpha_{11}&< {\cal O} (  10^{-14} ),\end{align} although the strong interaction and nuclear physics uncertainty in these bounds spans several orders of magnitude.
Neutron oscillations and nucleus decay    give much weaker constraints on coefficients of operators containing heavier quarks, as the required loops involve weak interactions and small CKM parameters. See, e.g., Ref.~\cite{Barbier:2004ez} for a summary of such constraints in R-parity--violating  supersymmetry.

We also need to check that  the new sources of flavor violation will not violate the stringent constraints coming from the agreement between theory and experiment in neutral meson oscillation phenomenology. After integrating out $\phi$ and the $N_i$, the interactions in Eq.~(\ref{eq:lag}) will lead to $\Delta F=2$ flavor-changing operators.The most stringent constraints come from the neutral kaon oscillations.

The relevant $\Delta S=2$ Hamiltonian that is generated reads~\cite{Gabbiani:1996hi}
\begin{align}
{\cal H}_{\phi N_i}^{\Delta S=2}=\sum_{i,j}\frac{y_{di}^\ast y_{dj} y_{si}y_{sj}^\ast}{2^9\, 3^3\, \pi^2 m_\phi^2} \bar d_R^\alpha\gamma^\mu s_R^\alpha\, \bar d_R^\beta\gamma_\mu s_R^\beta,
\label{eq:dS2}
\end{align}
where $\alpha$ and $\beta$ are color indices, and we have assumed that $m_\phi\simeq m_{N_i}$. This contributes to the $K^0$--$\bar K^0$ mass difference,
\begin{align}
\Delta m_K&=2{\rm Re}\bra{\bar K^0}{\cal H}_{\phi N_i}^{\Delta S=2}\ket{K^0}
\\
&={\rm Re}\sum_{i,j}\frac{y_{di}^\ast y_{dj} y_{si}y_{sj}^\ast}{2^8\, 3^4\, \pi^2}\frac{m_K f_K^2}{m_\phi^2},
\end{align}
with $f_K$ the kaon decay constant. Requiring that this not exceed the measured mass difference of $\left(3.484\pm0.006\right)\times10^{-12}~\MeV$~\cite{Agashe:2014kda} leads to a constraint on the couplings,
\begin{align}
\left({\rm Re}\sum_{i,j}y_{di}^\ast y_{dj} y_{si}y_{sj}^\ast\right)^{1/4}<0.40\sqrt{\frac{m_\phi}{650~\GeV}},
\end{align}
where we have used $f_K=155~\MeV$~\cite{Durr:2010hr,*Follana:2007uv}. Since the Yukawa couplings are complex, there is also a contribution to CP-violation in the neutral kaon system. This is typically parameterized by
\begin{align}
\epsilon_K&=\frac{{\rm Im}\bra{\bar K^0}{\cal H}^{\Delta S=2}\ket{K^0}}{\sqrt2 \Delta m_K}.
\end{align}
If we assume that the kaon mass difference is saturated by the SM contribution, then the operator in Eq.~(\ref{eq:dS2}) gives a contribution to CP-violation of
\begin{align}
\epsilon_K&=14\left(\frac{650~\GeV}{m_\phi}\right)^2{\rm Im}\sum_{i,j}y_{di}^\ast y_{dj} y_{si}y_{sj}^\ast.
\end{align}
Requiring that this is less than the measured value, $\left(2.228\pm0.011\right)\times10^{-3}$~\cite{Agashe:2014kda}, results in a  constraint,
\begin{align}
\left({\rm Im}\sum_{i,j}y_{di}^\ast y_{dj} y_{si}y_{sj}^\ast\right)^{1/4}<0.11\sqrt{\frac{m_\phi}{650~\GeV}}.
\end{align}

Since we expect that the CP-violating phases of the Yukawa couplings have no reason to be suppressed, the limit from $\epsilon_K$ is stronger than the one from $\Delta m_K$. However, this constraint does not impact the viability of the model; even ${\cal O}\left(1\right)$ values of $\left|y_{si}\right|$ are allowed so long as $\left|y_{di}\right|\lesssim {\cal O}\left(10^{-2}\right)$.

Limits from $B^0$-$\bar B^0$ and $B_s$-$\bar B_s$ mixing are no more constraining than from kaon mixing, and values of $\left|y_{bi}\right|$ as large as unity are allowed.

Experimental upper bounds on Electric Dipole Moments (EDMs) place strong constraints on new CP-violating physics. For new CP-violating physics at the weak scale, there may be nonstandard  contributions  to EDMs at one or two loops, which place constraints on combinations of the new couplings and phases.  The relevant constraints on R-parity--violating supersymmetry are summarized in Ref.~\cite{Barbier:2004ez}, and on supersymmetric models in general in Ref.~\cite{Pospelov:2005pr}.  In our minimal model, which does not include new couplings to leptons or to left handed quarks,  and which does not require any new couplings to be large, there are no one loop contributions or large two loop contributions to EDMs and no constraints on the CP-violating phases.

\section{Cosmology and the baryon-to-entropy ratio}
\label{sec:cosmology}
The baryon asymmetry of the Universe is quantified by the ratio of the net baryon number to entropy,
\begin{align}
\eta_B=\frac{n_B}{s}.
\end{align}
The most precise determination of this quantity comes from measurements of the CMB that fix the baryon density in units of the critical density, $\Omega_b$, which is related to the baryon-to-entropy ratio via $\eta_B = 3.9\times 10^{-9} \times \Omega_{b}h^{2}$. Planck has measured $\Omega_{b}h^{2} = 0.0221\pm0.0003$~\cite{Ade:2013zuv} which gives $\eta_B = \left(8.6\pm 0.1\right) \times 10^{-11}$.

In the minimal version of the model, the colored scalars can be produced at late times, well after they have frozen out, through $N_3$ decays. Those produced in decays after the universe has cooled below the QCD hadronization temperature $T_c\simeq 200~\MeV$ bind immediately to form mesinos and antimesinos. The CP-violating oscillations and decays (to antibaryons and baryons) of the (anti)mesinos give rise to a baryon asymmetry. To quantify this asymmetry, we must first determine the number of $N_3$ present after the universe has cooled below $T_c$.

There are two processes which control the number density of $N_{3}$: annihilation to quarks and scalars and decay to a scalar and quark. The number density of $N_3$ at time $t$ is then
\begin{align}
n_{N_3}=n_{N_3}^{\rm relic}e^{-\Gamma_{N_3}t}\left(\frac{a_{\rm relic}}{a_t}\right)^3.
\end{align}
Here, $n_{N_3}^{\rm relic}$ and $a_{\rm relic}$ are the number density of $N_{3}$ and the scale factor, respectively, at T $\approx$ $T\sim m_{N_3}/20$, and $a_t$ is the scale factor at $t$. Similarly to $N_{1}$ and $N_{2}$, we assume that $N_{3}$ is close in mass to $\phi$, $\Delta m_{\phi N_3}\ll m_\phi,m_{N_3}$. Then its decay rate is given by
\begin{equation}
\Gamma_{N_3} \simeq\sum_{q=d,s,b}\frac{y^{2}_{q3}}{8\pi}\frac{\Delta m^{2}_{\phi N_3}}{m_{\Phi_{q}}}.
\end{equation}
We want an appreciable number of $N_3$ to survive until $T_c\simeq 200~\MeV$ which corresponds to a time $t_c\sim 10^{-5}~\rm s$ assuming standard thermal history. Our thermal history is not standard since we have $N_{3}$ decaying and injecting energy into the plasma competing with the cooling of the plasma due to the expansion of the universe.  However, as we show below, that does not drastically affect the time vs. temperature relationship. Therefore, to get a respectable number of $N_3$ to decay after $T_c$, producing mesinos, we must have $\Gamma_{N_3}\lesssim 1/t_c\sim10^{-19}~\GeV$. For a splitting $\Delta m_{\phi N_3}\sim{\cal O}\left({\rm GeV}\right)$, this requires $y^{2}_{q3}\lesssim 10^{-15} (m_{N_{3}}/\TeV)$. Since the $N_3$ annihilation rate is proportional to $y^{4}_{q3}$, Yukawa couplings that are this small mean that the annihilation rate was always much smaller than the expansion rate of the Universe. Hence, assuming reheating happened at a high temperature and the $N_3$ were in equilibrium with the plasma, $n_{N_3}^{\rm relic}=\left(3/4\right)n_\gamma$ where $n_\gamma$ is evaluated at $T=m_{N_3}/20$ and only the decay rate $\Gamma_{N_3}$ dictates the $N_3$ number density for $T<T_c$.

We can determine the baryon-to-entropy ratio by co-evolving the $N_{3}$ and radiation energy densities,
\begin{align}
\frac{d\rho_{\rm rad}}{dt} &= -4 H \rho_{\rm rad} + \Gamma_{N_{3}} m_{N_{3}} n_{N_{3}},
\label{eq:drad}
\\
\frac{d\rho_{N_{3}}}{dt} &= -3 H \rho_{N_{3}} - \Gamma_{N_{3}} m_{N_{3}} n_{N_{3}},
\label{eq:dN3}
\end{align}
along with the baryon (minus antibaryon) number density,
\begin{align}
\frac{dn_B}{dt} &= -3 H n_B +\frac12 A\, \Gamma_{N_{3}} \epsilon_{B} n_{N_{3}}.
\label{eq:dnB}
\end{align}
Equations~(\ref{eq:drad}) and (\ref{eq:dN3}) describe the radiation and $N_3$ energy densities, taking into account the expansion of the universe and the fact that the decays of $N_{3}$ heat up the plasma. Equation~(\ref{eq:dnB}) describes the evolution of the net baryon number density, which develops a nonzero value due to CP-violating oscillations of mesinos produced by $N_{3}$ decays. The factor of $1/2$ appears due to the definition of $\epsilon_B$ in Eq.~(\ref{eq:eps}) as the baryon asymmetry per mesino-antimesino pair. $A$ counts the fraction of mesino states that undergo CP-violating oscillations. In this work, we focus on the strange mesino since, being an isoscalar, it is not decohered by scattering on the pions present in the plasma after confinement, a possibility in the case of the down mesino that would complicate the analysis. Thus, we take $A=1/3$, assuming that equal fractions of up, down, and strange mesinos are formed which should be true to $\sim 30\%$, the level of SU(3) flavor breaking observed elsewhere. It is important to keep in mind that in additon to Eq.~(\ref{eq:dnB}), the net baryon number density is subject to the constraint that $n_B=0$ for $T<T_c$ since scalars produced above the QCD confinement temperature do not hadronize and therefore do not undergo coherent oscillations.

Instead of solving Eqs.~(\ref{eq:drad}) to (\ref{eq:dnB}) exactly, we first use the sudden decay approximation to gain an intuitive understanding of the baryon-to-entropy ratio and to obtain simple analytic relations among the parameters of the model. In this approximation, we consider the comoving $N_3$ number density to be constant until the time $t_{\rm decay}=1/\Gamma_{N_3}$, after which it is zero. The $N_3$ decays dump energy into the plasma\footnote{Note that solving Eqs.~(\ref{eq:drad}) to (\ref{eq:dnB}) exactly, as we will do below, shows that the plasma temperature does not actually increase during $N_3$ decay---rather, its cooling rate slows~\cite{Scherrer:1984fd}. For this reason we ignore whether $N_3$ decays ``reheat'' the plasma to a temperature above $T_c$, a possibility that would naively spoil hadronization.} and produce scalars that, if the plasma temperature at the time of decay is less than $T_c$, form mesinos that oscillate and decay on time scales short compared to the expansion of the Universe,\footnote{At temperatures below $T_c$, the $\phi$-$\phi^\ast$ annihilation cross section becomes large, ${\cal O}(\Lambda_{\rm QCD}^2)$~\cite{Kang:2006yd}. However, the rate for this process is tiny compared to the oscillation and decay rates since it is suppressed by the small ratio of the $N_3$ to $\phi$ decay rates, $\Gamma_{\rm ann}\propto n_\phi\sim n_{N_3}\Gamma_{N_3}/\Gamma_\phi$.} creating a net baryon asymmetry. We now describe the evolution of the Universe in the context of this approximation.

At very early times, the Universe's energy density is dominated by radiation, until some time $t_{\rm eq}<t_{\rm decay}$ when the energy density in $N_3$ (matter) is equal to that in radiation. For times $t_{\rm eq}<t<t_{\rm decay}$, the Universe is matter-dominated. After the decays of $N_3$ at $t_{\rm decay}$, the Universe is again radiation dominated. The temperature of the plasma at the time of matter-radiation equality, $t_{\rm eq}$, is
\begin{align}
T_{\rm eq}&=\frac{45\zeta\left(3\right)}{2\pi^4\geff\left(T_i\right)}m_{N_3},
\end{align}
with $\geff\left(T_i\right)$ the effective number of relativistic degrees of freedom at a temperature $T_{i} \sim m_{N_3}$ when the universe is very hot and radiation dominated. We will use $\geff\left(T_i\right) = 116.25$ which takes into account the SM contributions as well as contributions from $\phi$ and $N_{1,2}$. The time vs. temperature relationship during radiation domination fixes the time at $T_{\rm eq}$,
\begin{align}
t_{\rm eq}&=\sqrt{\frac{45}{2\pi^2\geff\left(T_{\rm eq}\right)}}\frac{M_{\rm Pl}}{T_{\rm eq}^2}
\\
&=\left(\frac{2\pi^2}{45}\right)^{3/2}\left[\frac{\pi^2}{\zeta\left(3\right)}\right]^2\frac{\geff^2\left(T_i\right)}{\geff^{1/2}\left(T_{\rm eq}\right)}\frac{M_{\rm Pl}}{m_{N_3}^2},
\end{align}
where $M_{\rm Pl}=\left(8\pi G\right)^{-1/2}=2.4\times10^{18}~\GeV$ is the Planck mass and $\geff\left(T_{\rm eq}\right)\simeq70$, for $T_{\rm eq}$ of order a few $\GeV$. After $t_{\rm eq}$, assuming $t \gg t_{\rm eq}$ the $N_3$ and radiation energy densities evolve according to matter domination,
\begin{align}
\rho_{\rm rad}&=\frac{4M_{\rm Pl}^2}{3}\left(\frac{t_{\rm eq}}{t^4}\right)^{2/3},~\rho_{N_3}=\frac{4M_{\rm Pl}^2}{3}\frac{1}{t^2}.
\label{eq:rhos}
\end{align}
Defining $\xi\equiv \rho_{N_3}(t=t_{\rm decay}^-)/\rho_{\rm rad}(t=t_{\rm decay}^-)$, where $t_{\rm decay}^-$ indicates an infinitesimal time before decay, we find
\begin{align}
\xi&=\left(\frac{t_{\rm decay}}{t_{\rm eq}}\right)^{2/3}
\\
&=\left(\frac{45}{2\pi^2}\right)\left[\frac{\zeta\left(3\right)}{\pi^2}\right]^{4/3}\left(\frac{\geff\left(T_{\rm eq}\right)}{\geff^4\left(T_i\right)}\right)^{1/3}\left(\frac{m_{N_3}^2}{M_{\rm Pl}\Gamma_{N_3}}\right)^{2/3}.
\end{align}

The baryon-to-entropy ratio is
\begin{align}
\eta_B=\frac{n_B\left(t=t_{\rm decay}^+\right)}{s_{\rm rad}\left(t=t_{\rm decay}^+\right)},
\end{align}
where $t_{\rm decay}^+$ denotes an infinitesimal time after decay. The net baryon number density is related to the $N_3$ number density just before decay,
\begin{align}
n_B\left(t=t_{\rm decay}^+\right)=n_{N_3}\left(t=t_{\rm decay}^-\right)\times\frac12A\,\epsilon_B.
\end{align}
$A$ and $\epsilon_B$ are determined solely by the mesino particle physics and factor out of the cosmological story. We can write $\eta_B$ in terms of $\xi$,
\begin{align}
\eta_B=&\frac{\rho_{\rm rad}\left(t_{\rm decay}^-\right)}{s_{\rm rad}\left(t_{\rm decay}^+\right)}\frac{\xi}{m_{N_3}}\frac{A\epsilon_B}{2}
\\
=&\frac34\frac{T_{\rm rad}\left(t_{\rm decay}^-\right)}{m_{N_3}}\left[\frac{T_{\rm rad}\left(t_{\rm decay}^-\right)}{T_{\rm rad}\left(t_{\rm decay}^+\right)}\right]^3\xi\,\frac{A\epsilon_B}{2}.
\end{align}
The ratio of the temperatures before and after decay can be found by using the conservation of energy,
\begin{align}
\rho_{\rm rad}\left(t_{\rm decay}^+\right)&=\rho_{\rm rad}\left(t_{\rm decay}^-\right)+\rho_{N_3}\left(t_{\rm decay}^-\right)
\\
&=\rho_{\rm rad}\left(t_{\rm decay}^-\right)\left(1+\xi\right),
\end{align}
which leads to
\begin{align}
\eta_B=&\frac{3}{4}\frac{T_{\rm rad}\left(t_{\rm decay}^-\right)}{m_{N_3}}\frac{\xi}{\left(1+\xi\right)^{3/4}}\frac{A\epsilon_B}{2}.
\end{align}
We can use the expression for the radiation energy density in Eq.~(\ref{eq:rhos}) to find the temperature of the plasma just before decay, $T_{\rm rad}\left(t_{\rm decay}^-\right)\equiv T_{\rm decay}$,
\begin{align}
T_{\rm decay}\simeq&\frac{2}{\sqrt3}\left(\frac{\pi^2}{\zeta\left(3\right)}\right)^{1/3}\left[\frac{\geff\left(T_i\right)}{\geff\left(T_{\rm decay}\right)}\right]^{1/4}\left(\frac{M_{\rm Pl}^2\Gamma_{N_3}^2}{m_{N_3}}\right)^{1/3}
\\
\simeq&\frac{15\sqrt3\zeta\left(3\right)}{\pi^4}\frac{1}{\geff\left(T_i\right)}\frac{m_{N_3}}{\xi},
\end{align}
with $\geff\left(T_{\rm decay}\right) = 63.75$ the number of relativistic degrees of freedom at this temperature. We ignore here the decrease the number of degrees of freedom below the QCD confinement temperature since it obscures the physics and is numerically unimportant at the level of accuracy of our estimates. Using this expression, we can express the baryon-to-entropy ratio as
\begin{align}
\eta_B\simeq&\frac{45\sqrt{3}\zeta\left(3\right)}{8\pi^4}\frac{1}{\geff\left(T_i\right)}\frac{1}{\left(1+\xi\right)^{3/4}}A\epsilon_B
\\
=&6.1\times10^{-10}\left(\frac{116.25}{\geff\left(T_i\right)}\right)\left(\frac{10}{1+\xi}\right)^{3/4}
\\
&\quad\quad\times\left(\frac{A}{1/3}\right)\left(\frac{\epsilon_B}{10^{-5}}\right).
\end{align}
The factor
\begin{align}\left(1+\xi\right)^{-3/4}=\left[\frac{T_{\rm rad}\left(t_{\rm decay}^-\right)}{T_{\rm rad}\left(t_{\rm decay}^+\right)}\right]^3
\end{align}
can be thought of as the entropy dilution of the naive baryon-to-entropy ratio due to the $N_3$ decays heating the plasma.\footnote{Again, we stress that the plasma does not actually heat up in an exact treatment of Eqs.~(\ref{eq:drad}) to (\ref{eq:dnB}). See below.}

\begin{figure}
\includegraphics[width=\linewidth]{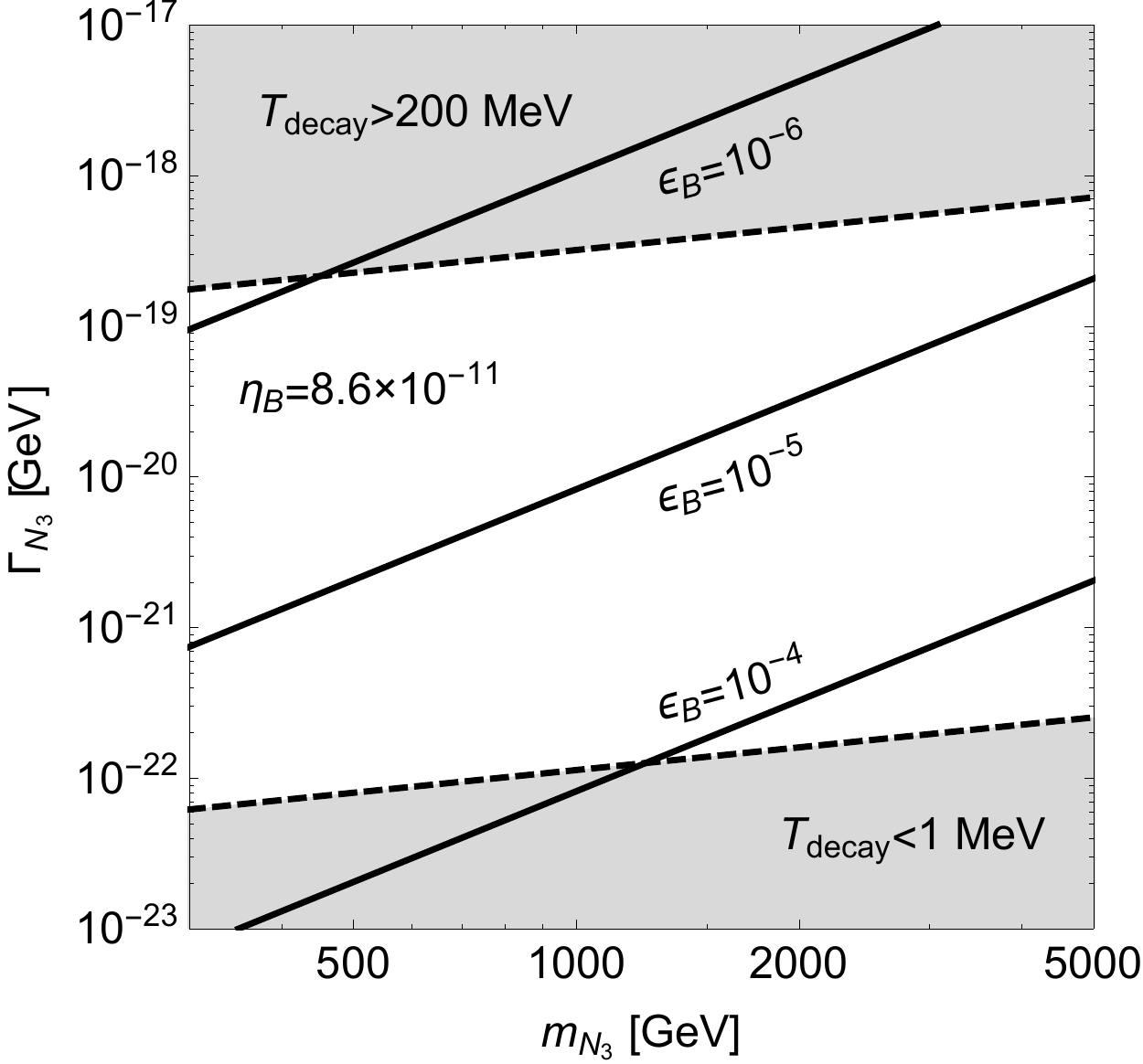}
\caption{Solid lines indicate values of $\Gamma_{N_3}$ and $m_{N_3}$ that give $\eta_B=8.6\times10^{-11}$ for $\epsilon_B=10^{-4},10^{-5}$, and $10^{-6}$, from bottom to top, respectively. The shaded regions lead to a plasma temperature at $N_3$ decay greater than $T_c\simeq200~\MeV$, in which case mesinos do not form, or a temperature at decay less than $1~\MeV$, which would spoil BBN. Note that this is calculated within the sudden decay approximation.}\label{fig:GammavsmN3}
\end{figure}

\begin{figure}
\includegraphics[width=\linewidth]{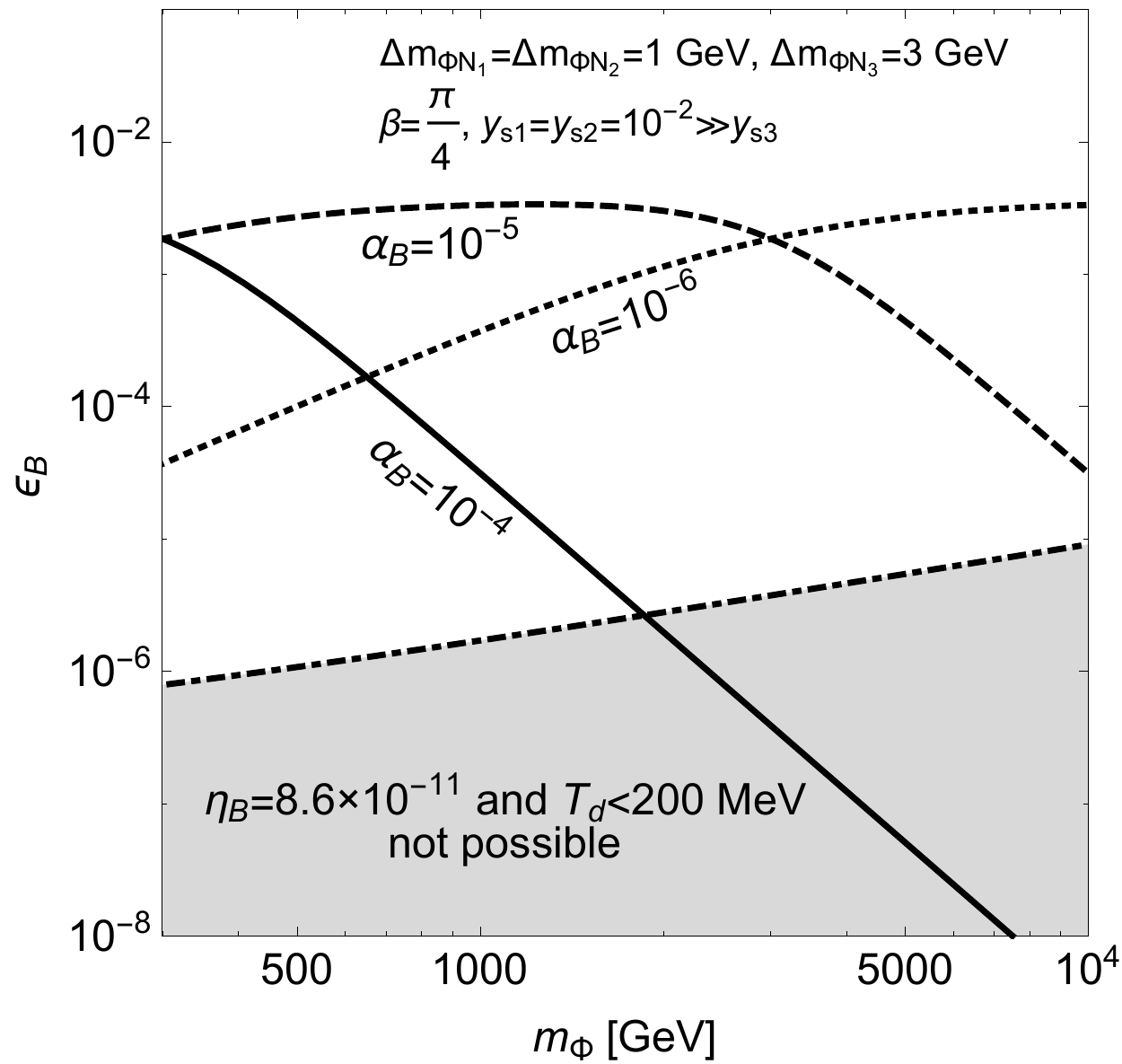}
\caption{The shaded region shows values of $\epsilon_B$ and $m_\phi$, assuming $\Delta_{\phi N_3}=3~\GeV$, that result in $\eta_B<8.6\times10^{-11}$ for $T_{\rm decay}<T_c\simeq200~\MeV$ in the sudden decay approximation. The points above this region can attain the measured value of $\eta_B$, depending on the value of $\Gamma_{N_3}$. Also shown are the values of $\epsilon_B$ when choosing $\Delta_{\phi N_1}=\Delta_{\phi N_2}=1~\GeV$, $y_{s1}=y_{s2}=10^{-2}\gg y_{s3}$, and $\alpha_{B}=10^{-4}$, $10^{-5}$, $10^{-6}$.}
\label{fig:epsvsmphi}
\end{figure}

For a fixed value of $m_{N_3}$, there is a minimal amount of entropy dilution, i.e. a minimal value of $\xi$, that can be found by setting the temperature of the plasma at $N_3$ decay to be equal to the QCD confinement temperature. Using this value of the entropy dilution determines a lower bound on the asymmetry $\epsilon_B$ as a function of $m_{N_3}$ such that a baryon-to-entropy ratio of $9\times10^{-11}$ is possible. We plot this lower bound on $\epsilon_B$ in Fig.~\ref{fig:epsvsmphi}, as a function of the scalar mass, taking the splitting between $N_3$ and $\phi$ to be $\Delta m_{\phi N_3}=3~\GeV$.

To assess whether we can reasonably achieve a baryon-to-entropy ratio in line with the current experimental value of $8.6\times10^{-11}$, we also show in Fig.~\ref{fig:epsvsmphi} the asymmetry $\epsilon_B$ as a function of $m_\phi$, taking $\Delta m_{\phi N_1}=\Delta m_{\phi N_2}=1~\GeV$, $y_{s1}=y_{s2}=10^{-2}\gg y_{s3}$, for $\alpha_B=10^{-4}$, $10^{-5}$, and $10^{-6}$. We see that there exist ranges of parameter choices that allow for scalars with a mass between $300~\GeV$ and $10~\TeV$ to have an asymmetry above the bound, and therefore to account for the measured value of $\eta_B$, depending on the choice of $\Gamma_{N_3}$. In principle, masses as large as $10^{6-7}~\GeV$ can be compatible with the observed baryon asymmetry but scales this large would require a larger $\epsilon_B$, hence a larger amount of fine-tuning. In addition, testing this model and embedding it in a framework that explains the weak scale, like the supersymmetric SM, could become difficult.

\begin{figure}
\includegraphics[width=\linewidth]{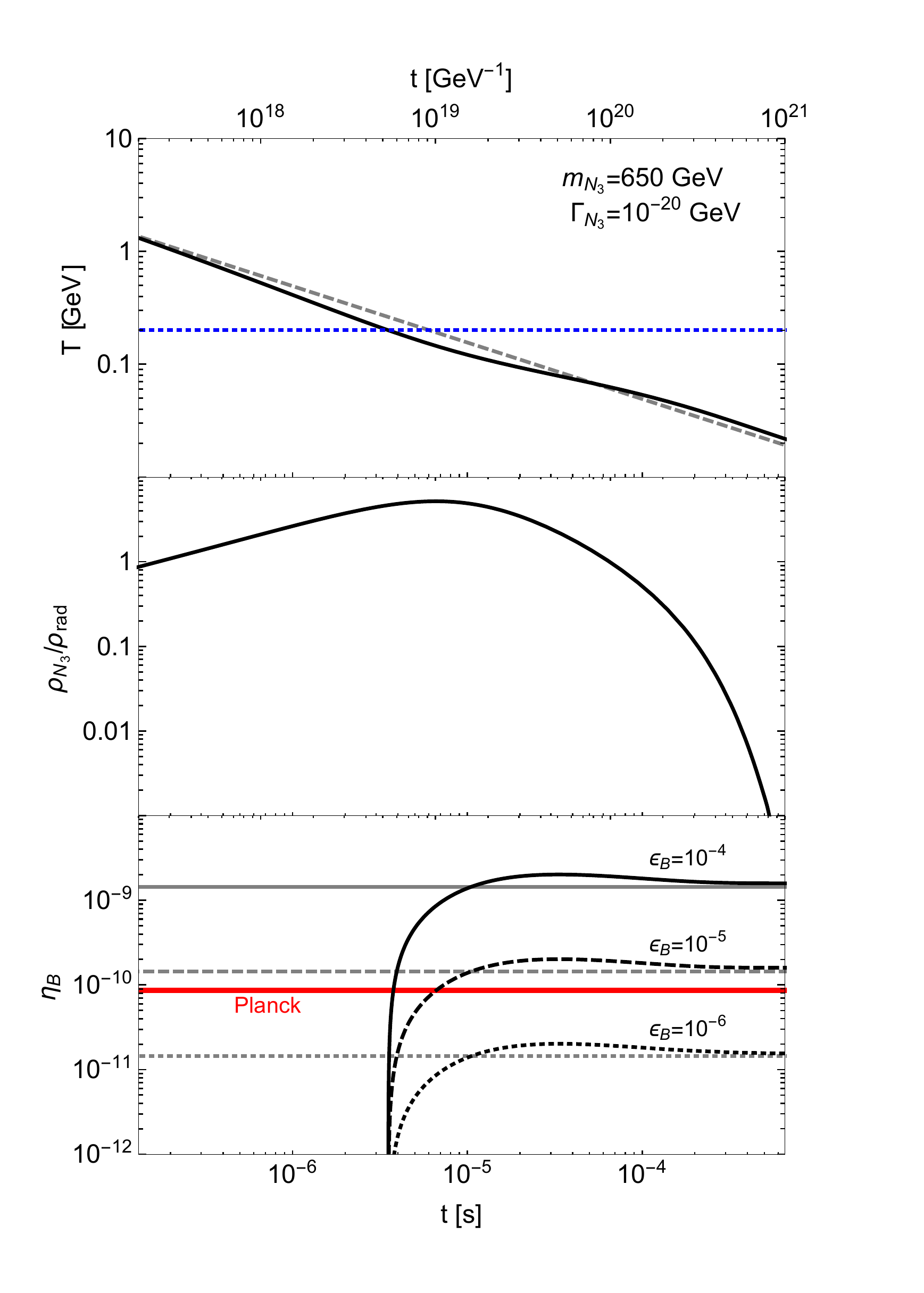}
\caption{Results of a numerical solution of Eqs.~(\ref{eq:drad}) to (\ref{eq:dnB}) for $m_{N_3}=650~\GeV$ and $\Gamma_{N_3}=10^{-20}~\GeV$. Top: the temperature of the plasma as a function of time (solid, black). For comparison, the evolution of the temperature without $N_3$ decays, assuming a radiation-dominated Universe is also shown (dashed, gray). The dotted blue line shows $T_c=200~\MeV$. As mentioned, the plasma does not actually heat up due to $N_3$ decays, rather it just cools more slowly. Middle: the evolution of the $N_3$ energy density to that of the plasma. Bottom: the growth of $\eta_B$ for values of the asymmetry $\epsilon_B=10^{-4}$ (solid, black), $10^{-5}$ (dashed, black), $10^{-6}$ (dotted, black) along with the value as measured by Planck (solid, red). Also shown are the sudden decay approximations to $\eta_B$ for each of these values of $\epsilon_B$ (gray). The decrease in $\eta_B$ from its maximum to its value at late times in each case reflects the entropy dilution due to the energy dumped into the plasma.}
\label{fig:numerics}
\end{figure}

We can check the accuracy of our sudden decay approximation by solving Eqs.~(\ref{eq:drad}) to (\ref{eq:dnB}) numerically and determining the resulting baryon-to-entropy ratio. In Fig.~\ref{fig:numerics}, we show the results of a numerical solution of the evolution equations with $m_{N_3}=650~\GeV$ and $\Gamma_{N_3}=10^{-20}~\GeV$. The top panel shows the evolution of the plasma temperature in time. As mentioned above, the plasma is not ``heated up'' by the decay of the $N_3$'s; instead, it cools more slowly. The next panel shows the ratio of the $N_3$ energy density to the radiation energy density, $\rho_{N_3}/\rho_{\rm rad}$, as a function of time. This ratio grows until $\sim 1/\Gamma_{N_3}$ when it decreases to to $N_3$ decays. The bottom panel shows the net baryon-to-entropy ratio, $\eta_B$ as it develops from zero at early times to a fixed value at late times for $\epsilon_B=10^{-4}$, $10^{-5}$, and $10^{-6}$. We also show the value of $\eta_B$ calculated for these parameters in the sudden decay approximation. We note that the sudden decay approximation of $\eta_B$ agrees with the numerical solution of Eqs.~(\ref{eq:drad}) to (\ref{eq:dnB}) to within an $\sim{\cal O}\left(1\right)$ factor for a wide range of values of $m_{N_3}$, $\Gamma_{N_3}$, and $\epsilon_B$.
In our numerical solution, we took $\geff\left(T\right)$ to be fixed at $63.75$ for temperatures less than $4~\GeV$. Including the change of $\geff$ at the hadronization temperature would not drastically affect our conclusions, and would only obscure the relevant physics.

We note here that the estimate of the baryon asymmetry in this section depended on the assumption that the scalars were produced by the late, out-of-equilibrium decay of a particle that was in thermal equilibrium at very high temperatures. Alternative cosmological histories are certainly possible. For example, the scalars could be produced by the decay of long-lived moduli~\cite{Banks:1993en}. Another possibility involves considering a nonthermal parent particle (possibly $N_3$) that dominates the energy density of the universe. Its decays to scalars reheat the universe and create a baryon asymmetry before the creation of a thermal plasma. Alternative scenarios may, of course, have a quite different relationship between $\epsilon_B$ and $\eta_B$ from the simple one presented here.

\section{Conclusions}
\label{sec:conclusions}
We have shown that  if there is a scalar quark which  lives long enough to hadronize, CP violation in the oscillations of mesinos is possible. If the scalar quark has baryon-number--violating decays, and if it can be produced out of thermal equilibrium in the early universe, such CP violation could be the origin of the matter-antimatter asymmetry of our universe. We presented a mechanism for the out-of-equilibrium production  of the scalar quark via the late decays of a very weakly coupled heavy neutral particle,  and computed $\eta_B$,  the ratio of  baryon number to entropy, in terms of the relevant parameters of the model. Because $\eta_B$ depends sensitively on the parameters,    the viable parameter space of the model is rather well constrained.   In particular, obtaining sufficient CP violation  requires the existence of  a neutral Majorana fermion which is not much lighter than the squark. Such a difficult to explain  coincidence hints at an environmental selection principle, which is attractive in light of the observation that the value of $\eta_B$ in our universe appears to be optimal for creating stars in galaxies  which are not so dense as to be unfriendly to advanced life~\cite{Tegmark:2005dy}.

In order to experimentally test the model one should begin by finding evidence for scalar quarks at the LHC. As discussed in Sec.~\ref{sec:colliders}, existing searches, especially in the case where the decays do not produce significant missing energy, are not very constraining. In our model the squark should  decay sometimes into  two antiquarks, and sometimes into a quark and a neutral fermion, with the latter decaying into one soft and two hard quarks. The final states are likely to contain at least one third generation quark for the reasons discussed in Sec.~\ref{sec:rare}. Because successful baryogenesis requires that the neutral fermion  not be much lighter than the mesino, the first quark is very soft. Thus  the scalar quark should  essentially decay into two hard jets (and possibly two very soft quarks). The decays may be prompt or could produce displaced vertices. The LHC reach for the discovery of heavy scalar quarks which decay into jets  is projected to reach 1070 GeV~\cite{Duggan:2013yna} with 1000 fb$^{-1}$. In Ref.~\cite{Berger:2012mm}, Berger, Csaki, Grossman, and Heidenreich proposed looking for same-sign top quarks as evidence for mesino oscillations, assuming that the squark predominantly decays into a top and a strange quark. In our model, same sign tops can be produced without mesino oscillations, as the neutral fermion which is often produced in $\phi$ decays is equally likely to decay to top and to antitop. Searching for such events containing same sign tops should increase the reach for searching for scalar quarks and  significantly constrain the model. Note that this signature will be CP-violating, but the small asymmetry between top and antitop will be too challenging to measure at the LHC. 

We also point out that a slight extension to the model would allow for asymmetric dark matter (for a review, see, e.g.,~\cite{Petraki:2013wwa,*Zurek:2013wia}) to be incorporated. If the operator mediating $\phi$ decays to baryons, $\alpha_{ij}\phi^\ast\bar d_i\bar u_j$, were modified to the dimension-5 operator $\alpha_{ij}\phi^\ast\bar d_i\bar u_j (\chi/M)$ where $\chi$ is a complex scalar and $M$ is the relevant scale, the model would possess a $\mathbb{Z}_2$ symmetry under which $\chi$, $\phi$, and $N_i$ are odd. This would render $\chi$ stable if it were the lightest of these particles. Then the oscillations and decays of the mesinos would produce a $\chi$-$\chi^\ast$ asymmetry along with the baryon-antibaryon asymmetry. The $\chi$ mass in this model would be fixed to be $\sim5~\GeV$ since the number densities of dark and baryonic matter would be equal. The signature of $\phi$ production at a hadron collider would now include missing energy, pushing up the lower bound on the allowed mass.
 
\appendix
\section{Form Factor Estimation}
\label{sec:Gamma12appendix}
In this appendix, we detail our estimation of the mesino-meson transition form factor which enters the two-body contribution to ${\bm \Gamma}_{12}$. We specialize to the $\Phi_s\to\eta N_1$ case that we considered above.

We begin by expressing the form factor using an operator product expansion (see, e.g., Ref.~\cite{Chernyak:1983ej} for a description of the formalism),
\begin{align}
F\left(q^2\right)&=\int\overline{dx}\,\overline{dy}\,\bra{\eta\left(p^\prime\right)}\bar s^i_\alpha s^k_\gamma\ket{0}C_{\alpha\beta\gamma\delta}^{ijkl}\bra{0}\phi^j\bar s^l_\delta\ket{{\Phi_s}_\beta\left(p\right)}.
\end{align}
$p$ is the mesino momentum and $p^\prime=p+q$ is the meson momentum---a transition involving an on-shell $N_1$ has $q^2=m_{n_1}^2$. $C_{\alpha\beta\gamma\delta}^{ijkl}$ is the short-distance Wilson coefficient calculated perturbatively. Greek subscripts are Lorentz indices and Latin superscripts are color indices. We use the shorthand $\int\overline{dx}=\int_0^1dx_1\int_0^1dx_2\delta(1-x_1-x_2)$ and similarly for $\overline{dy}$. $x_1$ is the momentum fraction of the squark in the mesino and $x_2$ is that of the light quark. $y_{1,2}$ are the momentum fractions of the quark and antiquark in the meson. We expand the bilinear operators in terms of operators of definite quantum numbers using the meson and mesino momentum distributions,\footnote{We ignore the $u\bar u$ and $d\bar d$ quark components of $\eta$, an approximation that is valid given the overall uncertainty of the calculation.}
\begin{align}
\bra{\eta\left(p^\prime\right)}\bar s^i_\alpha s^k_\gamma\ket{0}&=\frac{\delta^{ik}}{N_c}\left(-\frac{1}{4}\right)\left(\fslash{p}^\prime\gamma^5\right)_{\alpha\gamma}\phi^\ast_\eta\left(y_1,y_2\right)+\dots,
\\
\bra{0}\phi^j\bar s^l_\delta\ket{{\Phi_s}_\beta\left(p\right)}&=\frac{\delta^{jl}}{N_c}\delta_{\beta\delta}\phi_{\Phi_s}\left(x_1,x_2\right)+\dots~,
\end{align}
where the distribution amplitudes $\phi_\eta$ and $\phi_{\Phi_s}$ are normalized so that 
\begin{align}
\int\overline{dy}\phi_\eta\left(y_1,y_2\right)=f_\eta,~\int\overline{dx}\phi_{\Phi_s}\left(x_1,x_2\right)=f_{\Phi_s},
\end{align}
where $f_\eta$ and $f_{\Phi_s}$ are the meson and mesino decay constants.

\begin{figure}
\includegraphics[width=\linewidth]{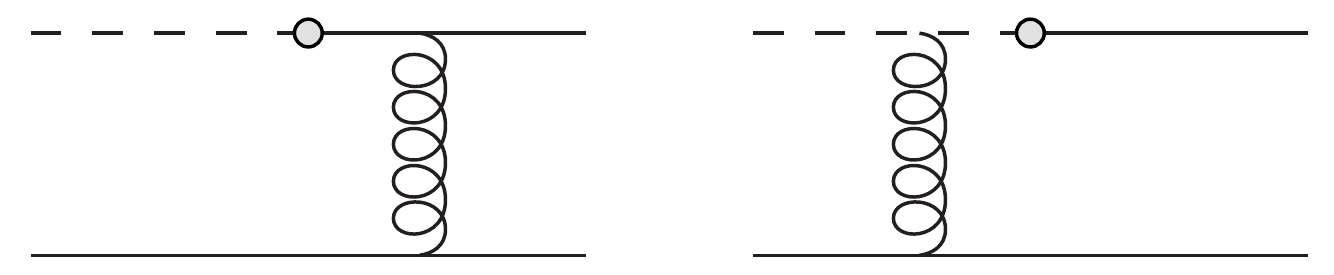}
\caption{Leading diagrams responsible for the $\Phi_s\to\eta N_1$ transition. Dashed lines are $\phi$ lines and solid lines are $s$ or $\bar s$ lines. Curly lines are gluons. Shaded dots represent the $\phi$-$s$-$N_1$ coupling.}\label{fig:diagrams}
\end{figure}

At leading order there are two diagrams that contribute to the Wilson coefficient, shown in Fig.~\ref{fig:diagrams}, which give
\begin{align}
C_{\alpha\beta\gamma\delta}^{ijkl}&=\left(ig_s\right)^2\left(-\frac{i}{k^2}\right)\left(\gamma_\rho\right)_{\gamma\delta} \left(t^a\right)^{ij}\left(t^a\right)^{kl}
\\
&\quad\times\left[\gamma^\rho \frac{i\!\not\!\!{\Delta}_1}{\Delta_1^2} P_L-\frac{i\left(x_1p+\Delta_2\right)^\rho}{\Delta_2^2-m_\phi^2} P_L\right]_{\alpha\beta} .
\nonumber
\end{align}
$\Delta_1=p^\prime-x_2p$ and $\Delta_2=p-y_2p^\prime$ are the momenta of the virtual quark and squark in the two diagrams and $k$ is the gluon momentum. Using these, the form factor becomes
\begin{align}
F\left(q^2\right)&=\pi\alpha_s\frac{{\rm tr}\left(t^at^a\right)}{N_c^2}
\\
&\quad\times\int\overline{dx}\,\overline{dy}\,\phi_M\left(x_1,x_2\right)\phi^\ast_P\left(y_1,y_2\right)\frac{\tilde C}{k^2}, \nonumber
\end{align}
with
\begin{align}
\tilde C&=\frac{1}{\Delta_1^2}{\rm Tr}\left(\fslash{p}^\prime\gamma^5\gamma^\rho\!\not\!\!{\Delta}_1 P_L \gamma_\rho\right)
\\
&\quad-\frac{1}{\Delta_2^2-m_\phi^2}{\rm Tr}\left(\fslash{p}^\prime\gamma^5 P_L \left(x_1\fslash{p}+\!\not\!\!{\Delta}_1\right)\right) \nonumber
\\
&=\frac{4p^\prime\cdot \Delta_1}{\Delta_1^2}+\frac{2p^\prime\cdot\left(x_1p+\Delta_1\right)}{\Delta_2^2-m_\phi^2}.
\end{align}
We will use simple forms for the distribution functions,
\begin{align}
\phi_\eta\left(y_1,y_2\right)&=6f_\eta\,y_1y_2,
\\
\phi_{\Phi_s}\left(x_1,x_2\right)&=f_{\Phi_s}\delta\left(x_2-\frac{\Lambda_{\rm QCD}}{m_{\Phi_s}}\right).
\end{align}
where $\Lambda_{\rm QCD}\sim300~\rm MeV$ is the QCD scale. The meson distribution function is motivated by the requirement that the quark and antiquark momentum distributions be symmetric---it can also be shown to be the asymptotic meson distribution amplitude. The mesino distribution function encodes the fact that the light quark carries a momentum fraction $m_q/m_{\Phi_s}$ where $m_q\sim\Lambda_{\rm QCD}$ is a constituent quark mass.

Integrating over the mesino momentum distribution takes
\begin{align}
\frac{4p^\prime\cdot \Delta_1}{\Delta_1^2}&\to -2+{\cal O}\left(\frac{\Lambda_{\rm QCD}}{m_{\Phi_s}}\right),
\\
\frac{2p^\prime\cdot\left(x_1p+\Delta_1\right)}{\Delta_2^2-m_\phi^2}&\to\frac{m_{\Phi_s}^2-q^2}{m_{\Phi_s}^2-m_\phi^2-y_2\left(m_{\Phi_s}^2-q^2\right)}
\label{eq:term2}
\\
&\quad+{\cal O}\left(\frac{\Lambda_{\rm QCD}}{m_{\Phi_s}}\right), \nonumber
\end{align}
and
\begin{align}
k^2&\simeq-\Lambda_{\rm QCD} m_{\Phi_s} y_2\left(1-\frac{q^2}{m_{\Phi_s}^2}\right) +{\cal O}\left(\frac{\Lambda^2}{m_{\Phi_s}^2}\right).
\end{align}
Because of the factor of $1/k^2$ in the Wilson coefficient, the form factor is enhanced in the region where $q^2$ approaches $m_{\Phi_s}^2$ giving the form factor a $1/\left(m_{\Phi_s}^2-q^2\right)$ scaling. After integrating over the meson momentum distribution, (\ref{eq:term2}) gives a logarithmic correction to the scaling which we can ignore. We therefore take $\tilde C\simeq-2$, finding
\begin{align}
F\left(q^2\right)&\simeq \frac{3\left(N_c^2-1\right)}{N_c^2}\pi\alpha_s\frac{f_\eta f_{\Phi_s}}{\Lambda_{\rm QCD} m_{\Phi_s}}\frac{1}{1-q^2/m_{\Phi_s}^2}
\\
&=\frac{8\pi\alpha_s}{3}\frac{f_\eta f_{\Phi_s}}{\Lambda_{\rm QCD} m_{\Phi_s}}\frac{1}{1-q^2/m_{\Phi_s}^2}.
\end{align}
Using $\alpha_s\simeq0.3$, appropriate for the mass splitting between $\Phi_s$ and $N_1$ of $\sim1~\GeV$ that we have in mind, $\Lambda_{\rm QCD} \simeq300~\MeV$, a value of the $\eta$ decay constant derived from the $\eta\to\gamma\gamma$ rate ($f_\eta=130~\MeV$), and the mesino decay constant from Eq.~(\ref{eq:decayconst}), this gives, for the production of a physical $N_1$,
\begin{align}
F\left(m_{N_1}^2\right)&\simeq 1.2\times10^{-2}\sqrt{\frac{650~\GeV}{m_{\Phi_s}}}\left(\frac{1~\GeV}{\Delta m_{\Phi N_1}}\right).
\end{align}

Being an estimate of an exclusive process in QCD, this calculation has considerable uncertainty,  and represents the largest contribution to the overall uncertainty in our estimate of the baryon-to-entropy ratio. However, it captures the essential physics: this rate is enhanced when the singlet mass is close to the mass of the mesino. This is because the quark produced in the $\phi\to N_1 s$ splitting has a smaller momentum in this regime, allowing it to more easily bind with the spectator quark to form a meson. Similar behavior is observed in the $B$ meson system, where the semileptonic transition form factor in $B\to\pi\ell\nu$ decays grows as the $\ell\nu$ mass approaches $m_B$.

Note that the overall normalization of the form factor, for a fixed $\left|{\bm \Gamma}_{12}\right|$, is   degenerate with the value of $\left|y_{s1}\right|$. 

\bibliography{baryogenesis}

\end{document}